\documentclass{tMOP2e}
\usepackage{amsmath,psfrag,graphics}

\begin{document}
\title{Tailoring PDC speckle structure.}
\author{G.Brida$^{a}$, M.Genovese$^{a}$$^{\ast}$\thanks{$^\ast$Corresponding
author. Email: m.genovese@inrim.it \vspace{6pt}}, A.Meda$^{a}$, E. Predazzi$^{b}$ and I.Ruo-Berchera$^{a}$  \\\vspace{6pt}
\vspace{3pt} $^{a}${\em{Istituto Nazionale di Ricerca Metrologica, I-10135 Torino, Italia}}\\
\vspace{3pt} $^{b}${\em{Dipartimento di Fisica Teorica Universit\`a di Torino and INFN, I-10125 Torino, Italia}}\\
\vspace{3pt}\received{xx/xx/xxxx}} \maketitle
\begin{abstract}
Speckle structure of parametric down conversion light has recently
received a large attention due  to relevance in view of
applications to quantum imaging

The possibility of tailoring the speckle size by acting on the pump
properties is an interesting tool for the applications to quantum
imaging and in particular to the detection of weak object under
shot-noise limit.

Here we present a systematic detailed experimental study of the
speckle structure produced in type II PDC with particular attention
to its variation with pump beam  properties.

\bigskip

\maketitle
\end{abstract}

\section{Introduction}

Thermal or pseudothermal light (as the one obtained by scattering of
coherent light by a diffuser) presents a random intensity
distribution known as speckle pattern \cite{las}. This structure can
have interesting applications, e.g. in metrology \cite{uto}.

In particular, speckle structure of parametric down conversion
(PDC) light has recently received a large attention due  to
relevance in view of applications to quantum imaging \cite{qi}.

The aim of Sub Shot Noise (SSN) quantum imaging is to obtain the
image of a weak absorbing object with a level of noise below the
minimum threshold that is unavoidable in the classical framework
of light detection. Being interested in measuring an image, one is
forced to consider a multi-mode source, which is able to display
quantum correlation also in the spatial domain. Theoretically,
this goal can be achieved by exploiting the quantum correlation in
the photon number between symmetrical modes of SPDC . Typically
the far field emission is collected by a high quantum efficiency
CCD camera. It is fundamental to set the dimension of the modes
coherence areas with respect to the dimension of the pixels. In
particular, the single pixel dimension must be of the same order
of magnitude of the coherence area or bigger, in order to fulfill
the sub-shot noise correlation condition, compatibly with the
requirement of large photon number operation. Thus, the
possibility of tailoring the speckle size by acting on the
intensity and size of the pump beam represents an interesting tool
for the applications to quantum imaging and in particular to the
detection of weak objects under shot-noise limit
\cite{qi,lug2,dit,and}.

A detailed theory of correlations and speckle structure in PDC has
been developed in \cite{lug} and, in another regime, in \cite{mat}
\footnote{see also \cite{mat2} for the seeded case.}Furthermore,
experimental results were presented in \cite{dit}.

Nevertheless, a systematic comparison of experimental variation of
speckle size and their correlations with the theoretical results
of \cite{lug} is still missing.

In this paper we present a systematic detailed experimental study
of the speckle structure produced in type II PDC with particular
attention to its variation with pump beam properties. In
particular dependence on pump power and size are investigated in
detail: results that will represent a test bench for the
theoretical models.

\section{Theory}\label{theory}

The process of SPDC is particularly suitable for studying the
spatial quantum correlations \cite{MG}, because it takes place with
a large bandwidth in the spatial frequency domain. Any pair of
transverse modes of the radiation (usually dubbed idler and signal),
characterized by two opposite transverse momenta $\mathbf{q}$ and
$-\mathbf{q}$, are correlated in the photon number, i.e. they
contain, in an ideal situation, the same number of photons. In the
far field zone, the single transverse mode is characterized by a
coherence area, namely the uncertainty on the emission angle
$\vartheta$ ($\tan\vartheta=\lambda q/2\pi$, $\lambda$ being the
wavelength) of the twin photons. It derives from two effects that
participate in the relaxation of the phase matching condition. On
the one side the finite transverse dimension of the gain area inside the crystal,
coinciding with the pump radius $w_{p}$ at low parametric gain. On
the other side the finite longitudinal dimension of the system, i.e.
along the pump propagation direction, that is generally given by the
crystal length $l$.

\begin{figure}[tbp]
\begin{center}
\includegraphics[ width=\textwidth, bb=0 0 1000 800]{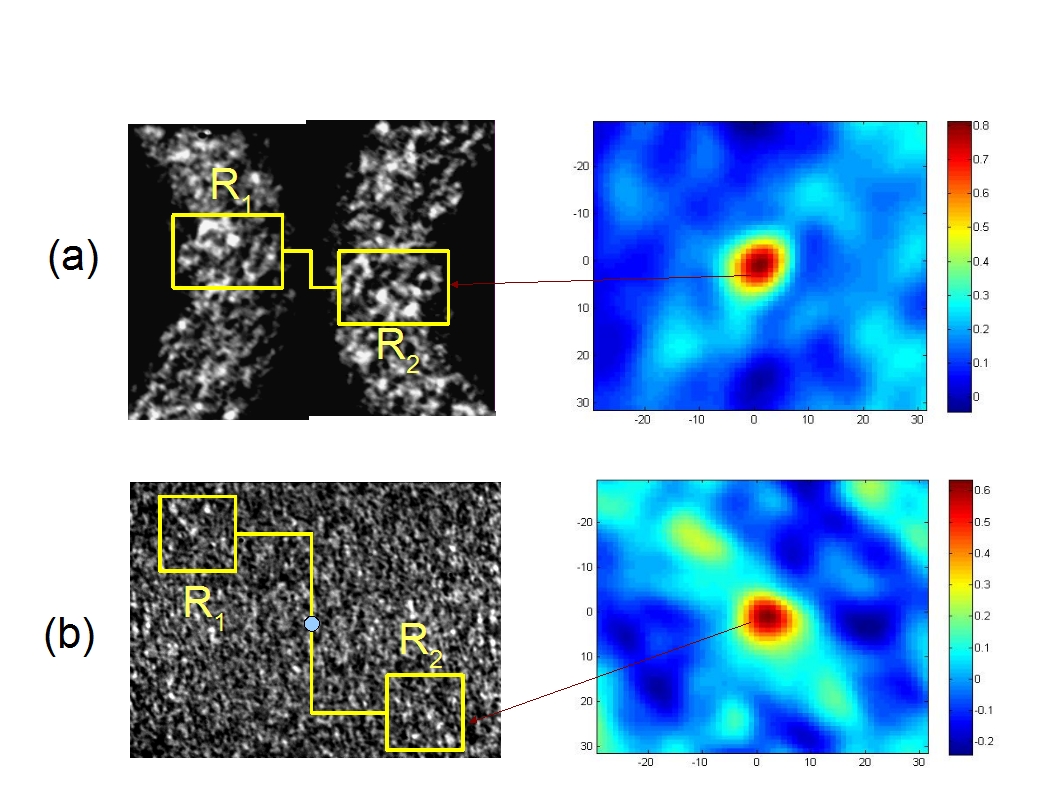}
\caption{CCD's images of the speckled structure of the type II PDC emission from a
BBO non-linear crystal. The two partial rings in (a) represent the two correlated beams 1 and 2,
selected by a narrow filter (Bandwidth=10 nm) around the degeneracy ($\omega_{1}=\omega_{2}=\omega_{p}/2$),
whereas in (b) no filtering is applied. On the right hand site, are shown the respective cross-correlation functions,
evaluated by fixing a region $R_{1}$ in
the signal field and moving a region $R_{2}$ in the idler field. A point in the cross-correlation graph
correspond to the  correlation coefficient between $R_{1}$ and $R_{2}$ for a single position of $R_{2}$.
The axes report the pixel by pixel displacement of region $R_{2}$.}\label{cross-corr}
\end{center}
\end{figure}

The appearance of the emission is a speckled structure in which
the speckles have, roughly, the dimension of the coherence area
and for any speckle at position $\mathbf{q}$ there exists a
symmetrical one in $-\mathbf{q}$ with equal intensity. This is
rather evident in the ccd images of SPDC shown in Fig.
\ref{cross-corr}.

In the following we summarize, very briefly, some elements of the
theory describing this structure.

Omitting some unessential constants, the Hamiltonian describing the
three fields parametric interaction is

\begin{equation}\label{QH}
\widehat{H}_{I}(t)\propto\int_{\mathcal{V}}\chi^{(2)}\;\widehat{\mathbf{E}}^{(+)}_{1}(\mathbf{r},t)\;
\widehat{\mathbf{E}}^{(+)}_{2}(\mathbf{r},t)\;\widehat{\mathbf{E}}^{(-)}_{p}(\mathbf{r},t)\;d^{3}r
+ h.c
\end{equation}

The evolution of the quantum system guided by Hamiltonian
(\ref{QH}), in the case of relatively high gain regime and
non-plane-wave pump, requires a numerical solution and it is
discussed in detail in \cite{lug}. Anyway, in the low gain regime
an analytical solution is achievable \cite{kl,bu}. Therefore, it
is worth to briefly mention the result in the first order of the
perturbation theory ($g\ll1$) for a gaussian pump, where the
quantum state of the scattered light has the entangled form

\begin{eqnarray}
\left| \psi \right\rangle &=&\left| \mathrm{vac} \right\rangle+\exp
\left[ -\frac{i}{\hbar }\int \widehat{H}_{I}dt
\right] \left| 0\right\rangle\nonumber\\
&=& \left| \mathrm{vac}
\right\rangle+\sum_{\mathbf{q}_{1},\mathbf{q}_{2}}
\sum_{\Omega}F(\mathbf{q}_{1},\mathbf{q}_{2},\Omega)\left|
1_{\mathbf{q}_{1},\Omega}\right\rangle \left|
1_{\mathbf{q}_{2},-\Omega}\right\rangle,
\end{eqnarray}

\begin{eqnarray}\label{F}
F(\mathbf{q}_{1},\mathbf{q}_{2},\Omega)&=&g\cdot
\,\mathrm{sinc}\left[\frac{\Delta k
(\mathbf{q}_{1},\mathbf{q}_{2},\Omega)\cdot
l}{2}\right]e^{-(\mathbf{q}_{1}+\mathbf{q}_{2})^{2}\frac{w_{p}^{2}}{4}},\nonumber\\
&&\omega_{1}= \omega_{p}/2+\Omega,
\qquad\omega_{2}=\omega_{p}/2-\Omega.
\end{eqnarray}

$\mathbf{q}_{1,2}$ being the transverse wave vectors and
$\omega_{p}, \omega_{1}, \omega_{2}$ the pump, idler and signal
frequencies respectively. The coherence area, in the limit of low
parametric gain $g$, can be estimated by the angular structure of
the coincidence probability
$\left|F(\mathbf{q}_{1},\mathbf{q}_{2})\right|^{2}$ at some fixed
frequency $\Omega$. As mentioned before, now is clear that we deal
with two functions that enter in the shaping of the coherence
area: the $sinc$ function and the Fourier transformed gaussian
pump profile. Since they are multiplied, the narrower determines
the dimension of the area.  The Half Width Half Maximum of the
gaussian function, appearing in (\ref{F}), is $\delta q=
\sqrt{2\ln( 2)}/w_{p}$. If we expand the longitudinal wave
detuning around the exact matching point $\Delta
k(\mathbf{q}_{0},\mathbf{q}_{0},\Omega)$, the linear part
\cite{bu} dominates for angles $\vartheta_{0}(\mathbf{q}_{0})$,
not too next to collinear regime and the $sinc$ function turns out
to have a HWHM of $\Delta q = 2,78/(l\tan\vartheta_{0})$ at
degeneracy. On the other hand, around the collinear emission, the
quadratic term prevails \cite{lug} and the bandwidth becomes
$\Delta q = 2,78*(2\pi/\lambda l)^{1/2}$. Concerning our
experiment, we consider small (but not zero) emission angles
$\vartheta$ and large enough pump radius $w_{p}$, such that we
always work in the region $\delta q/\Delta q< 1$. Therefore, in
principle, the dimension of the coherence area is only determined
by the pump waist.

When moving to higher gain regime, which is of greater interest for
our experiment,  the number of photon pairs generated in the single
mode increases exponentially as $\propto \sinh^{2}(g)$ i.e. a large
number of photons is emitted in the coherence time along the
direction $\vartheta$. In this case, also the pump amplitude becomes
important in the determination of the speckles dimension. As
described in \cite{lug2},  this can be explained by a qualitative
argumentation: inside the crystal, the cascading effect that causes
the exponential growth of the number of generated photons is
enhanced in the region where the pump field takes its highest value,
i.e. close to the center of the beam. Thus, in high gain regime,
most of the photon pairs are produced where the pump field is closed
to its peak value. As a result the effective region of amplification
inside the crystal becomes narrower than the beam profile. Thus, in
the far field one should consider the speckles as the Fourier
transform of the effective gain profile, that being narrower,
produces larger speckles.

A further fundamental consideration for the practical implementation
is that in high gain regime, instead of measuring the coincidences
between two photons by means of two single photo-detectors, one
collects a large portion of the emission by using for instance a CCD
array with a certain fixed exposure time. Within this time several
photons are collected by the single pixels and the result is an
intensity pattern, having the spatial resolution of the pixel. Looking at the images,
we can appreciate the speckled structure and a certain level of correlation of the speckles intensity between the
signal and idler arms (Fig. \ref{cross-corr}(a) and (b)). We can define also the
auto-correlation function of the signal intensity pattern itself,
since in the single transverse mode of the signal arm there are many
photons. To be precise the speckle's dimension is better related to
the spread of this function, although the two functions, the cross-
and the auto- correlation, present in the very high gain regime the
same behaviour with respect to the pump parameters, see \cite{lug}.

From the experimental view-point it is convenient to study the
auto-correlation because of the higher
visibility that allows a more accurate estimation of its size.

\section{Experiment}

\begin{figure}
\begin{center}
\includegraphics[width=15 cm, height=10 cm, bb=0 0 1200 900]{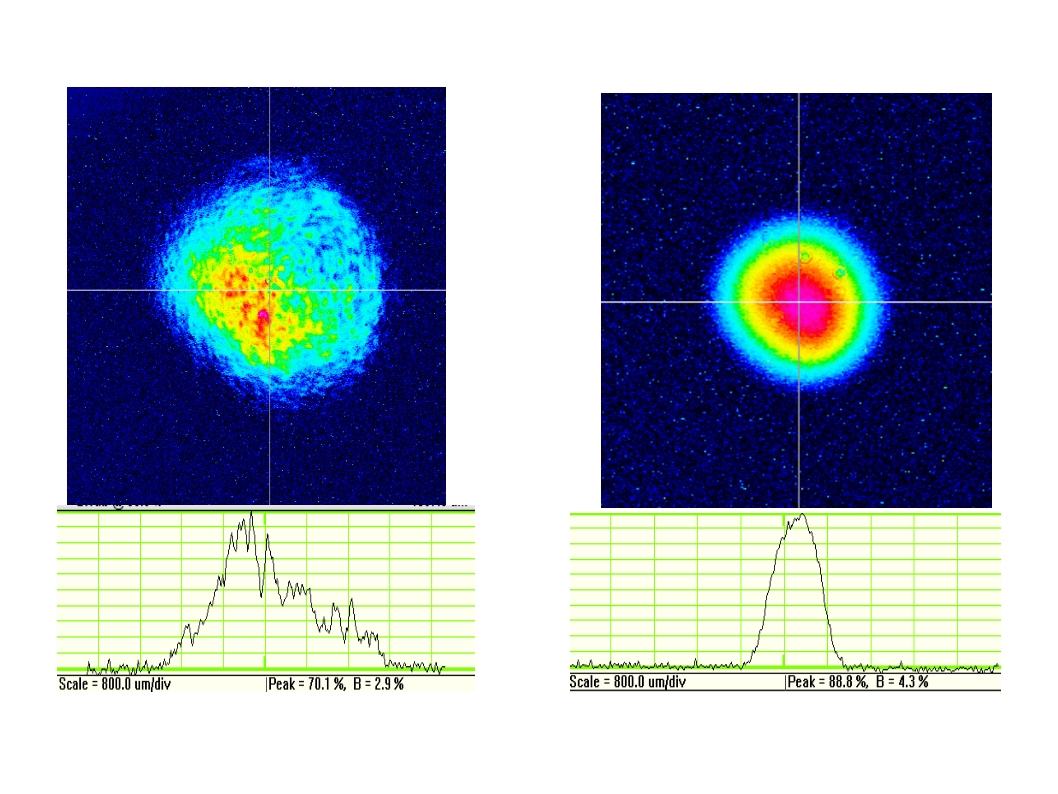}
\caption{Image of the pump beam before (on the left) and after (on the right) spatial filtering.}\label{pump}
\end{center}
\end{figure}

\begin{figure}[tbp]
\begin{center}
\includegraphics[angle=0, width=12 cm, height=8 cm]{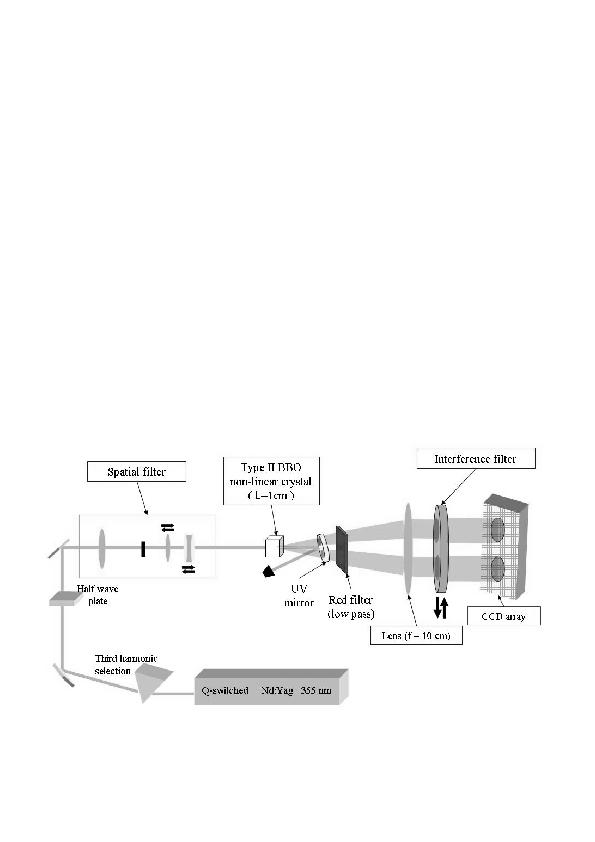}
\caption{Experimental setup. A triplicated Nd-Yag laser beam, after
spatial filtering, produces type II PDC in a BBO crystal, which is
then measured, after an interference filter and pump elimination, by
a CCD camera. }\label{fig2}
\end{center}
\end{figure}

Our setup is depicted in Fig. \ref{fig2}. A type II BBO non-linear
crystal ($l=1$ cm) is pumped by the third harmonic (wavelength of
355 nm ) of a Q-switched Nd:Yag laser. The pulses have a duration
of 5ns with a repetition rate of 10 Hz and a maximum energy, at
the selected wavelength, of about 200 mJ. The pump beam crosses a
spatial filter (a lens with $f$=50 cm and an iris of 250 $\mu$m of
diameter), in order to eliminate the non-gaussian components (see
fig. \ref{pump}). Before entering the crystal the pump is
collimated and its diameter is varied, when necessary, by changing
the distance between two lenses (a biconvex and a biconcave)
placed after the spatial filter. After the crystal, the pump is
stopped by a UV mirror, transparent to the visible (T=87\%
measured at 685nm), and by a low frequency-pass filter. The down
converted photons (signal and idler) pass through a lens of 5 cm
of diameter ($f=10$ cm) and an interference filter centered at the
degeneracy $\lambda$=710 nm (10nm bandwidth) and finally measured
by a CCD camera. We used a $1340X400$ CCD array, Princeton
Pixis:400BR (pixel size of 20 $\mu$m), with high quantum
efficiency (80\%) and low noise ($\Delta=4$ electrons/pixel). The
far field is observed at the focal plane of the lens in a $f-f$
optical configuration, that ensures that we image the Fourier
transform of the crystal exit surface. Therefore a single
transverse wavevector $\mathbf{q}$ is associated to a single point
$\mathbf{x}=(\lambda f/2\pi)\mathbf{q}$ in the detection plane.
The CCD acquisition time is set to 90 ms, so that each frame
corresponds to the PDC generated by a single shot of the laser.

Let us define $N_{R}(\mathbf{x})$ the intensity level,
proportional to the number of photons, registered by the pixel in
the position $\mathbf{x}$ of a region $R$. $\delta
N_{R}(\mathbf{x})= N_{R}(\mathbf{x})-\langle
N_{R}(\mathbf{x})\rangle$ is the fluctuation around the mean value
that is estimated as $\langle
N_{R}(\mathbf{x})\rangle=(1/n)\sum_{\mathbf{x}}
N_{R}(\mathbf{x})$, with $n$ the number of pixels. We evaluate the
normalized spatial auto-correlations of the intensity fluctuations
by choosing a large arbitrary region $R$, belonging for instance
to the signal portion of the image, and measuring

\begin{equation}\label{auto-corr}
 C(\mathbf{\xi})=\frac{\left\langle\delta
 N_{R}(\mathbf{x})\delta N_{R}(\mathbf{x+\xi})\right\rangle}
 {\sqrt{\left\langle \delta N_{R}(\mathbf{x})^{ 2}\right\rangle\left\langle\delta
 N_{R}(\mathbf{x+\xi})^{2}\right\rangle}}.
\end{equation}
where  $\xi$ is the displacement vector in the pixel space and
assumes discrete values. $C(\mathbf{\xi})$ has unit value for
$\mathbf{\xi}=0$.

\begin{figure}
\begin{center}
\includegraphics[width=15 cm, height=10 cm, bb=0 0 1000 800]{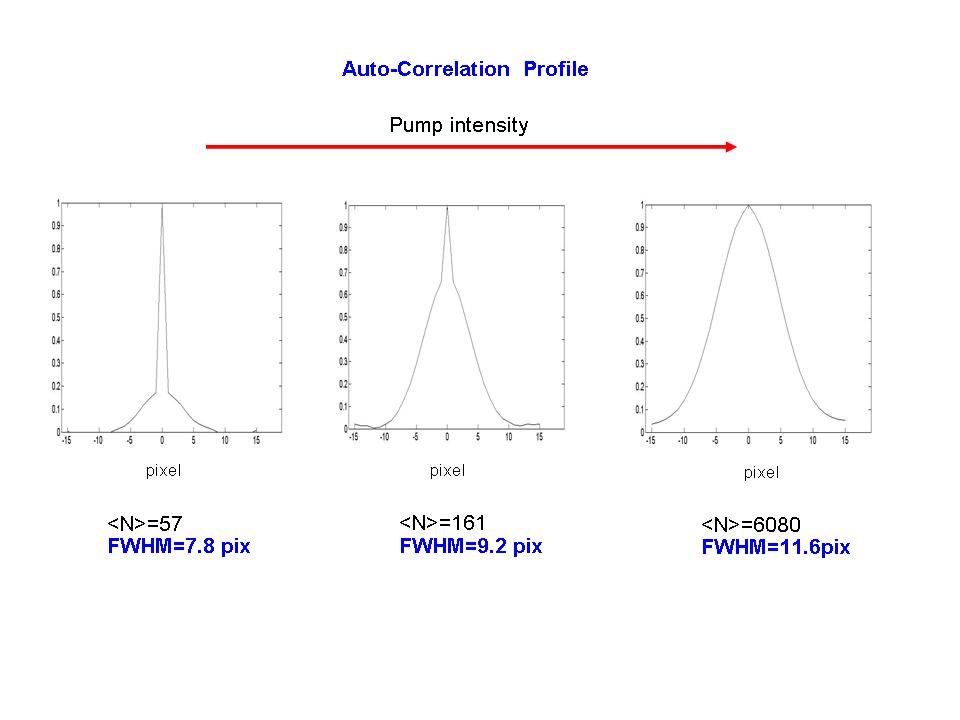}
\caption{Section of the two dimensional auto-correlation function evaluated in single shot images at different pump intensity}\label{auto-corr}
\end{center}
\end{figure}
\begin{figure}
\begin{center}
\includegraphics[width=15 cm, height=10 cm, bb=0 0 1000 700]{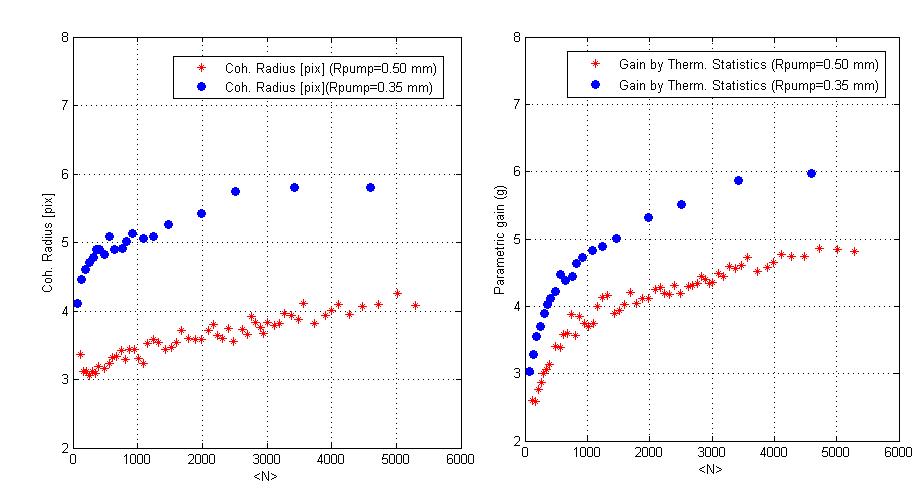}
\caption{On the left, speckle's radius and, on the right,
parametric gain estimation, both as function of photon number per
pixel, for two pump radii }\label{g2g5r2r5}
\end{center}
\end{figure}

Fig. \ref{auto-corr} shows some typical section of the
auto-correlation for different value of the pump intensity. One
note that, especially for small photon number, the
auto-correlation is dominated by the peak at $\mathbf{\xi}=0$ and
drop down even for displacement of one pixel, creating
symmetrically a sort of shoulder. This is explained by the
presence of noise at the scale of the pixel: first of all the
shot-noise that gives a contribution of $\langle
N_{R}(\mathbf{x})\rangle$, then the electronic noise in the
read-out and digitalization process is $\Delta$ and, finally, the
quantum efficiency that fluctuates with a standard deviation
$\delta\eta=3\%$ (we evaluated it using a self-calibration
procedure similar to the one described in \cite{Jiang}). On the
other side the fluctuations at the scale of the coherence area of
the light are given by  $\langle N_{R}(\mathbf{x})\rangle^{2}/M$,
where $M$ is the number of  modes (spatial modes $*$ temporal
modes), collected by one pixel. Therefore a large number of
spatial modes limits the visibility of the genuine
auto-correlation of the field with respect to the other sources of
noise due to the detection by CCD. Anyway we can take the width of
the shoulder as the most correct estimation of the coherence
radius. Fig. \ref{g2g5r2r5} reports on the left hand side, one of
our main results, namely the speckle's radius $R_{coh}$ as a
function of the photon number per pixel for two different pump
size $w_{p}=1.0$ mm and $w_{p}=0.7$ mm. The increasing of the
radius of the auto-correlation function versus the mean number of
photons can not be explained by the theory in the low gain regime,
predicting no dependence on the gain, i.e. on the pump intensity.
This is a signature of the fact that we are working in a non
linear regime of PDC in which the parametric gain $g$ does not
fulfill the condition $g\ll 1$, as we will discuss in the next
paragraph.
It must be noticed that we are constrained in the range of
intensities of the pump. For high intensities we are limited by
the damage threshold of optical components, for low intensities by
the visibility of the speckle structure because we collect a lot
of temporal modes in the same frame.

In order to comment the behavior of the radius it is important to
evaluate the effective gain region. We adopt two approaches. First
we try to evaluate $g$ by the statistics of the light.
Approximatively, according to Fig. \ref{g2g5r2r5}, we can consider
the dimension of the pixel always smaller than the coherence area
and thus the number of collected spatial modes $\sim 1$
\cite{goodman}. We evaluate the number of temporal modes $M$,
within this approximation, by using the property of the thermal
statistics corrected for the presence of experimental
imperfections described above:

\begin{equation}\label{thermal}
\langle \delta^{2}N\rangle=\langle N\rangle+\frac{\langle N\rangle^{2}}{M}+\delta^{2}\eta\langle N\rangle^{2}+\Delta^{2}
\end{equation}
The estimation is performed on a statistical ensemble of
independent pixel belonging to a single shot frame, thus is
definitely a spatial statistics.

After that, we can approximatively estimate the parametric gain by
the relation
\begin{equation}\label{gain}
\langle N\rangle=\eta\cdot\eta_{coll}\cdot M\cdot\sinh^{2}(g),
\end{equation}
where  $\sinh^{2}(g)$ is the number of photons in the single
space-temporal mode, $\eta$ is the quantum efficiency (including
losses) and $\eta_{coll}$ is the collection efficiency. It takes
into account that the pixel collects only a portion of the spatial
mode and roughly is given by the ratio between the pixel area and
the auto-correlation area $\eta_{coll}\approx A_{pix}/A_{coh}$.
The result is reported in Fig. \ref{g2g5r2r5}, on the right hand
side. As expected, the gain $g$ is larger than $1$, confirming
that we are working in high gain regime. Let also notice that,
although each point is an average over tens of images they are
distributed rather noisy, as well as the radius (on the left
insert).

\begin{figure}
\begin{center}
\includegraphics[width=\textwidth, bb=0 0 1000 700]{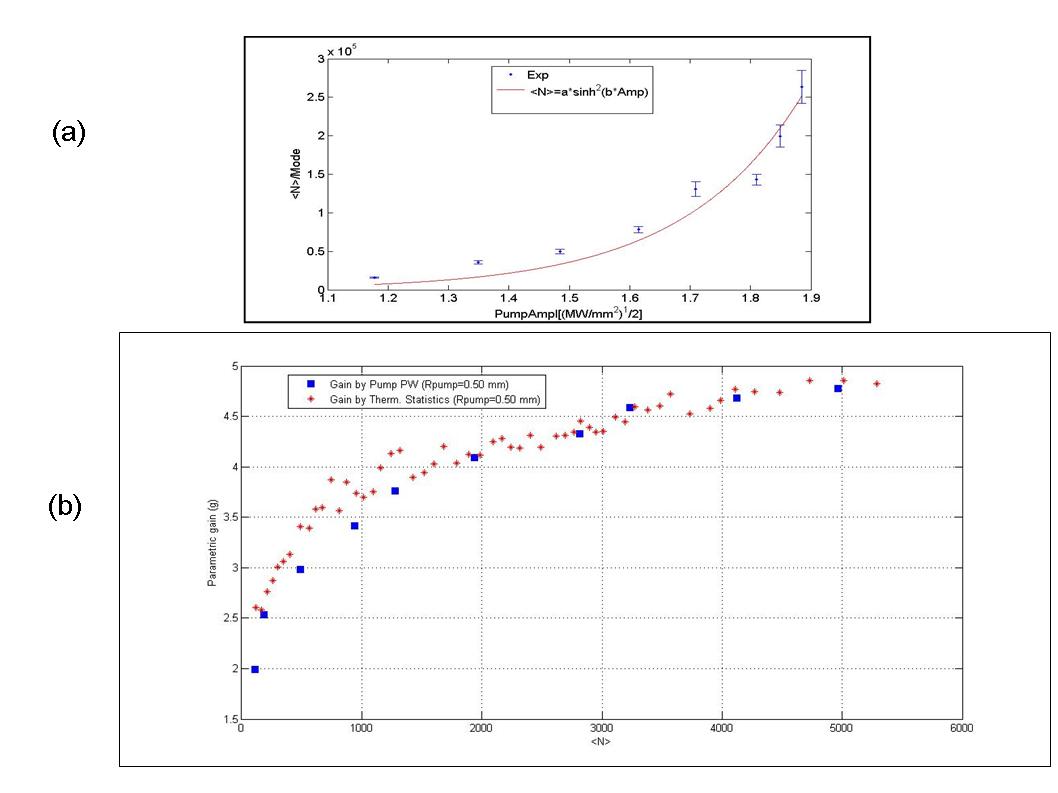}
\caption{(a) Number of photons in a single mode as function of
pump amplitude  (b) Parametric gain as function of the photon
number per pixel estimated from (a) and from the property of the
thermal statistics}\label{GainvsN}
\end{center}
\end{figure}

In order to explain the main features of the data it is
fundamental to take into account the characteristics of the pump.
It is the third harmonic of a multimode high power laser and the
power is varied by the delay between the Q-switch turn-on and the
lamp flash. The calibration curve delay-power (mean pulse power)
has been measured by a power meter, and we observed a
reproducibility with uncertainty around 10\%. The fluctuation of
the mean pulse power is about 20\% (measured by a photodiode).
Since the mean number of photons is proportional to $
\sinh^{2}(g)$, the fluctuations of the pump power generate large
fluctuations in the photons number from frame-to-frame.

Furthermore, a multimode laser when the number of modes $n$ is
larger than few units should present a thermal statistics of the
intensity fluctuation according to the formula $\delta
I/I=\sqrt{1-1/n}$ \cite{goodman} and, after the non linear process
of second and third harmonic generation the fluctuations can
increase even more. The expected temporal profile of the output
pulse is very far from a smooth gaussian function. Therefore,
taken into account the strong non-linearity of the PDC generation,
it is reasonable to consider that, inside the single pulse
entering the crystal, only few peaks of intensity contribute to
the large majority of the PDC emission. The main consequence is
that the number of \textit{effective} temporal modes $M$,
collected in the single shot, is much less than what expected by
evaluating the ratio between the pulse duration (5 ns) and the PDC
coherence time $\tau_{p}\sim 1$ ps and $M$ also fluctuates
randomly from image to image. By using Eq. \ref{thermal}, we
obtain a mean number of effective temporal modes
$\overline{M}=170\pm80$ when $w_{p}=1.0$ mm and
$\overline{M}=60\pm40$ for $w_{p}=0.7$ mm. At the same time, the
values of the parametric gain $g$ are in general bigger than 1
(see Fig \ref{g2g5r2r5}). The gain region in which we are working
ensures the possibility to reach a sufficient non-linear cascading
effect in the photon pairs production inside the crystal.

As a confirmation of the high non-linearity obtained in our
experiment we estimate the gain $g$ in an independent way \cite{masha}. In Fig.
\ref {GainvsN}(a) we present the measurement of the number of
photon in the spatial mode defined as $\langle N_{mode}\rangle=
\langle N\rangle\cdot A_{coh}$ by varying the mean pump amplitude
$A_{pump}$, while the pump size is fixed to $w_{p}=1.0$ mm. Here
the mean pump amplitude is the square root of the average power of
the single pulse divide by the transverse pump area measured by a
CCD. The data are fitted by the equation $\langle
N_{R}\rangle=k\cdot\sinh^{2}(\sigma A_{pump})$, where $k$ is fixed
to $k=\eta \overline{M}$, while $\sigma$ ($\sigma A_{pump}=g$) is the free
parameter. The experimental values, mediated on three different
acquisitions, are $\sigma=2.53\pm0.04 [\mathrm{mm^{2}/W}]^{1/2}$.
Fig. \ref {GainvsN}(b) shows the comparison between the values of
the gain obtained in this way (green squares) and in the previous
statistical approach (the red dots). The data are very close, and
we can conclude that we are actually in a strong non linear
regime.
\begin{figure}[tbp]
\begin{center}
\includegraphics[width=15 cm, height=10 cm, bb=0 0 1000 600]{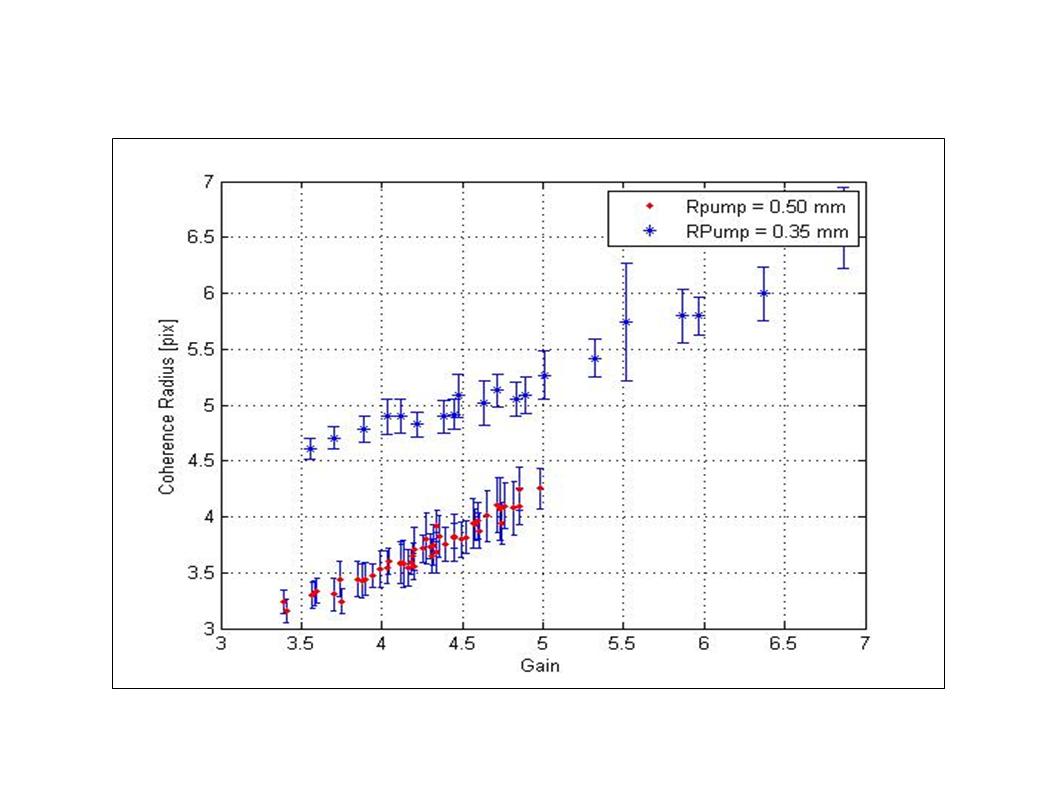}
\caption{Coherence radius as function of the parametric
gain}\label{r2r5}
\end{center}
\end{figure}

Now, let us consider the trend of the coherence radius versus the
parametric gain for the two different pump size as reported in
Fig. \ref{r2r5}. The data are well fitted by two straight lines
(not showed in figure), $R_{i}(g)=\alpha_{i} g+R_{i}(0)$ with
$i=1,2$. We obtain that $\alpha_{1}(0.67 \pm 0.06)$ and
$\alpha_{2}(0.55 \pm 0.06)$ agree within the their uncertainty.
Although a comparison  with a theoretical model in high gain
regime is needed for a proper interpretation of our data, we can
consider this behavior as an evidence of the influence of the gain
and of the pump diameter on the dimension of the speckles. In
fact, for the same value of the gain, the speckles radius is
always larger when the pump size is smaller as predicted in the
low gain.
\begin{figure}[tbp]
\begin{center}
\includegraphics[width=15 cm, height=10 cm, bb=0 0 1000 700]{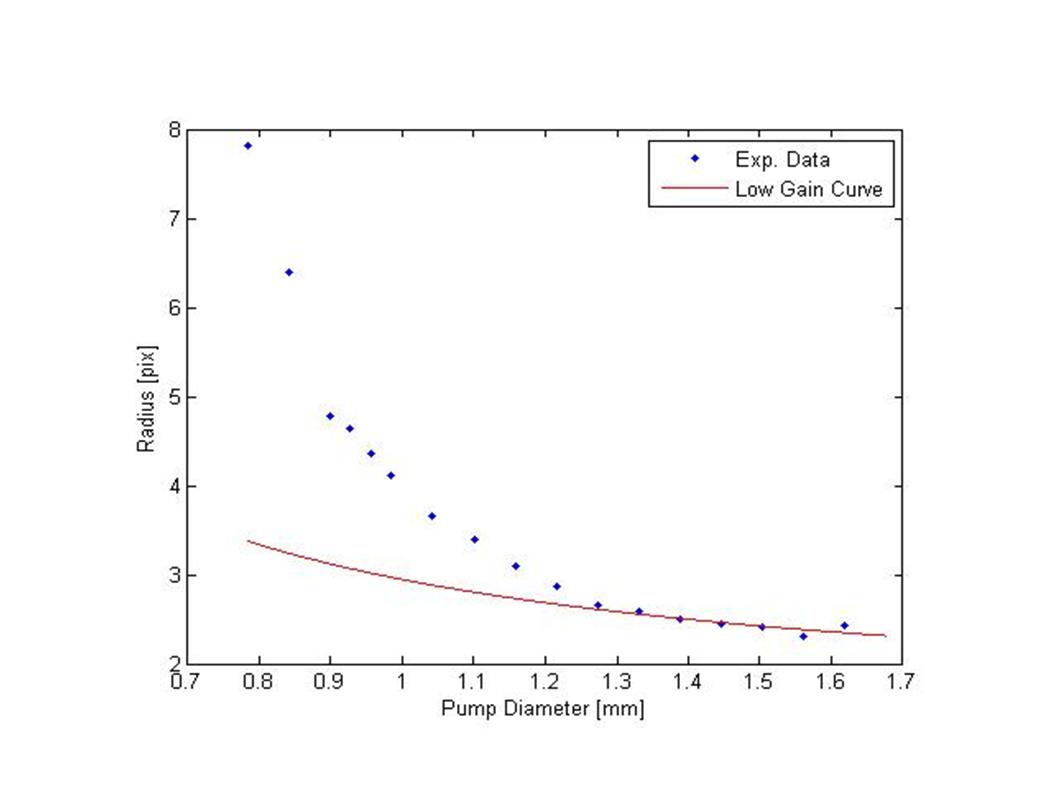}
\caption{Observed dependence of the radius (in pixels) of the
speckles from the pump diameter }\label{Pdiam_R}
\end{center}
\end{figure}

Finally, we investigate the dependence of the radius of the
speckles from the pump diameter, as shown in figure \ref{Pdiam_R}.
We fix the power of the laser to 0,78 MW, within the fluctuations,
and we change the diameter varying the distance between the two
collimating lenses (see figure \ref{fig2}). Since the diameter
changes, the intensity and the gain $g$ change as well. The
theory, in low gain regime, provides that the radius of the
speckles is proportional to the inverse of the pump size ($w_{p}$)
obtaining the red line in figure. It approach the experimental
data only when the pump diameter is large. This confirms the role
of the high gain regime in the speckle size. In fact, together
with the reduction of the pump diameter, the gain increases, and
thus the effective gain area is even more reduced. This effect
impresses upon the speckles size a stronger dependence with
respect to the pump size.

\section{Conclusion}

This paper provides a detailed experimental study of the size of
the coherence area in PDC in the high gain regime. We show that
the speckles present, not only a dependence on the pump diameter,
as in the usual low gain regime, but also a strong dependence form
the pump intensity. The understanding of the behavior of the
coherence area in the high gain regime is fundamental for the
innovative application in the field of quantum imaging. Our
results provide the basis for a comparison with the theoretical
models; in general, the observed dependencies of speckles size on
pump parameters allow the tailoring of speckles properties, an
instrument useful not only for applications to quantum imaging but
also for quantum metrology, quantum information, etc.

\subsection{ Acknowledgments}
This work has been supported by MIUR (PRIN 2005023443-002), by
07-02-91581-ASP, by Regione Piemonte (E14) and by Fondazione
Compagnia di San Paolo. Thanks are due to Alessandra Gatti, Enrico
Brambilla, Lucia Caspani, Luigi Lugiato and Ottavia Jedrkiewicz
for useful discussions.

\label{lastpage}

\end{document}